\documentstyle[11pt]{article}

\newcommand{\be}{\begin{equation}}
\newcommand{\ee}{\end{equation}}
\newcommand{\ba}{\begin{array}}
\newcommand{\ea}{\end{array}}
\newcommand{\bc}{\begin{center}}
\newcommand{\ec}{\end{center}}
\newcommand{\bi}{\begin{itemize}}
\newcommand{\ei}{\end{itemize}}

\newcommand{\disregard}[1]{{}}

\def\bild#1\over#2{\mathrel{\mathop{\kern0pt #1}\limits_{#2}}}

\begin{document}
\centerline {{\bf   ANYONIC PARTITION FUNCTIONS }}

\centerline{{\bf and}}

\centerline{{\bf WINDINGS OF PLANAR BROWNIAN MOTION\rm}}
\vskip 2cm
\centerline{{\bf Jean DESBOIS, Christine HEINEMANN\rm } and
{\bf St\a'ephane OUVRY \rm }\footnote{\it  and
LPTPE, Tour 12, 3\a`eme \a'etage, Universit\'e Paris  6, 75005 Paris
Cedex / electronic e-mail:
OUVRY@FRCPN11}}
\vskip 1cm
\centerline{{Division de Physique Th\'eorique \footnote{\it Unit\a'e de
Recherche  des
Universit\a'es Paris 11 et Paris 6 associ\a'ee au CNRS},  IPN,
  Orsay Fr-91406}}
\vskip 3cm
{\bf Abstract :  The computation of the $N$-cycle brownian
paths contribution $F_N(\alpha)$ to the $N$-anyon partition function is
adressed. A detailed
numerical analysis based on random walk on  a lattice indicates that
$F_N^{(0)}(\alpha)= \prod_{k=1}^{N-1}(1-{N\over k}\alpha)$.
In the paramount $3$-anyon case, one can show that
$F_3(\alpha)$ is  built by linear states belonging to the bosonic,
fermionic, and mixed representations of $S_3$.}

\vskip 1cm

IPNO/TH 94-55

June 1994

\vfill\eject

In this note one considers 2-dimensional identical particles
with fractional statistics (anyons) [1]. This system seems to
play an
important role in the understanding of the Fractional Quantum Hall Effect [2].

The $N$-anyon
Hamiltonian (there is no additional mutual interaction) reads
\begin{equation}
H_N={1\over
2m}\sum_{I=1}^N(\vec p_I -\alpha\vec A_I)^2+\sum_{I=1}^N{1\over
2}m\omega^2
{\vec  r_I}^2
\end{equation}
where $\alpha$ is the statistical parameter ($\alpha=0$ corresponds to the
bosonic
case, $\alpha=1$ to the fermionic case), and $\vec A_I$ the statistical
vector potential. An harmonic potential term has been added
as a long distance regulator. It discretizes
 the energy spectrum, and allows for a convenient computation of
thermodynamical quantities in the thermodynamic limit ($\omega\to 0$,
 where $\omega$ is the harmonic oscillator
pulsation).

In a recent series of papers [3], the $N$-anyon partition function
has been shown to be written as
\begin{equation}
Z_N(\alpha)=\sum_{\cal P}F_{\cal
P}(\alpha)\prod_L{1\over\nu_L!}\left({Z_1(L\beta)\over
L}\right)^{\nu_L}
\end{equation}
where summation is understood over all partitions ${\cal P}$ of $N$.
Each partition  $\sum_L L\nu_L = N$
is in
one-to-one correspondence with  an equivalence class of the permutation
group
$S_N$, namely the class of permutations which can be written as a product of
cycles of
length $L$, $\nu_L$  times.
The one-particle partition
function $Z_1(\beta)= {1/ (2\sinh{\xi\over 2})^2}$
is obviously statistics independant ( $\xi=\hbar\beta\omega,
\beta={1\over kT}$). Note that in (2), the statistical
information has been completely factorized out in the
$F_{\cal P}(\alpha)$'s, which satisfy
\be F_{\cal P}(1+\alpha)=sgn({\cal P})F_{\cal P}(\alpha)\ee
It follows that $Z_N(\alpha)$ is naturally written as
a sum of symmetric and
antisymmetric terms under $\alpha\longrightarrow 1+\alpha$.
By convention one has $F_{\cal P}(0)=1$ for all ${\cal P}$, i.e. $\alpha=0$
corresponds to Bose statistics.

In a brownian motion language, the
$F_{\cal P}$'s are simply expressed as
\begin{equation} F_{\cal
P}(\alpha)=<e^{i\alpha\sum_{I<J}\theta_{IJ}}>_{({\cal C})} \end{equation}
where the average
$<\;>_{(\cal C)}$
is taken over all possible brownian paths which induce, in the configuration
space,   a given
permutation ${\cal  P}$, i.e. which correspond to a given partition  of
the integer $N$. Note that these curves are closed
curves in the configuration space quotiented by the permutation group
$S_N$.  The  $\theta_{IJ}$'s are the relative angles between any pair of
particles $\{I,J\}$ (see below for a more precise definition).
For example, $F_N$ is
associated with cycles of
maximal length $L=N$, with $\nu_L=1$. Such cycles, involving the $N$ particles
of the system,
are by definition    $N$-particles paths over a time interval
$\tau=\hbar\beta$. Equivalently, they can be considered as  $1$-particle
paths over a time interval  $N\tau$, simply because the particles are
identical
[3]. Therefore, $<\;>_{(\cal C)}$ can be  taken over all brownian
paths of length $N\tau$. These paths are divided into $N$ segments of
equal
length $\tau$, each segment corresponding to the motion of a point $M_I$ (all
the
$M_I$'s have the same speed). The $\theta_{IJ}$'s are defined as the
angles of the vectors ${\vec {M_IM_J}}$ with a fixed axis in the plane
($-\infty<\theta_{IJ}<+\infty$) (see fig. 1 for $F_3$).

In the simplest $N=2$ case, eq. (2), has been exactly solved  [1].
It is instructive to display the results and their intimate connection
with probability distributions of planar brownian windings.
$Z_2(\alpha)$ is given as
\begin{equation}
Z_2(\alpha) ={1\over 2} F_2(\alpha)Z_1(2\beta)+{1\over
2}F_{11}(\alpha)Z_1(\beta)^2
\end{equation}
and is computed from the complete linear states
\be \psi_{n,m}(r,\theta)=e^{im\theta} e^{-\beta { r^2\over 2}}
r^{|m-\alpha|} L_n^{|m-\alpha|}(\beta r^2)\ee
\be E_{n,m}=\omega(2n+|m-\alpha|+1)\ee
where the $L_n$'s are the Laguerre polynomials.
One gets for $F_2(\alpha)$ and $F_{11}(\alpha)$
\be {F_2(\alpha) ={\sinh(1-2\alpha){\xi\over 2}\over  \sinh{\xi\over
2}}\qquad
F_{11}(\alpha) ={\cosh(1-2\alpha){\xi\over 2}\over  \cosh{\xi\over
2}}}\ee
In terms of brownian winding, the Fourier transform of
$F_{11}(\alpha)$ $(0<\alpha<1)$ yields the probability for a brownian curve to
wind
an even number of half-windings  around a fixed point, i.e. the usual result
for an integer brownian winding [4]. $F_2(\alpha)$ leads to the probability for
an odd number of half-windings [3].

The second virial coefficient follows directly [5]
\be a_2(\alpha)=
{1\over
 4}-{(1-\alpha)^2\over 2}\ee
One notes that in the thermodynamic limit where
the harmonic potential vanishes
\begin{equation}
F_2(\alpha) \longrightarrow_{\omega\rightarrow
0} F_2^{(0)}(\alpha) =1-2\alpha\ee

\be F_{11}(\alpha) \longrightarrow_{\omega\rightarrow 0}
F_{11}^{(0)}(\alpha) =1
\ee
The trivial thermodynamical limit of $F_{11}(\alpha)$
is easily understandable
in terms of (4), since in this limit there is no chance for the $2$
independent $1$-particle brownian paths to overlap.

In the  $N=3$ case, things become non trivial, since only a part of the
spectrum (the linear states [6] that generalize (6)) is known. One has by
definition
\begin{equation}
Z_3(\alpha) =
{1\over 3}F_3(\alpha)Z_1(3\beta)+{1\over
2}F_{21}(\alpha)Z_1(\beta)Z_1(2\beta)+{1\over
6}F_{111}(\alpha)Z_1(\beta)^3
\end{equation}
It happens that the  antisymmetric part of $Z_3(\alpha)$, namely
$F_{21}(\alpha)$, is known [7] to be exactly built from the linear states, i.e.
\be Z_1(\beta)
Z_1(2\beta) F_{21}(\alpha)=Z_3^{lin}(\alpha)-Z_3^{lin}(\alpha+1)\ee
with [6]
\be
Z_3^{lin}(\alpha)={\cosh[3(1-\alpha)\xi]\over
32\sinh^2{\xi\over 2}\sinh^2\xi \sinh^2{ 3\xi\over 2}}\ee
This yields
\be F_{21}(\alpha) ={\sinh(1-2\alpha){3\xi\over 2}\over  \sinh{3\xi\over
2}}\ee
with
\be F_{21}(\alpha)\longrightarrow_{\omega\rightarrow
0}F_{2}^{(0)}(\alpha)\ee
as it should.

In [3], numerical estimations for
$F_3^{(0)}(\alpha)$, and for $F_{111}(\alpha)$ at order $\xi^4$, have
been performed in the thermodynamic limit
\begin{equation}
F_3^{(0)}(\alpha)\sim (1-3\alpha)(1-{3\over 2}\alpha)
\end{equation}
This numerical result, altogether with $F_{111}^{(4)}(\alpha)$, is
sufficient  to propose a
simple analytical expression for the third virial coefficient [3]
\be a_3(\alpha)\sim
{1\over
36}+{1\over 12\pi^2}\sin^2(\pi\alpha)\ee

Having reached this point, and looking at the simple expression for
$F_3^{(0)}(\alpha)$,
it seems natural to suspect
that it could be generalized to $N>3$, and that it might be
derived from pure theoretic considerations. This is what is now going to
be shown.

Indeed, using (4), one is tempted to make numeric simulations on
a lattice for the $F_{\cal P}(\alpha)$'s. Note that in the thermodynamic
limit, one has always that the $F_{\cal P}^{(0)}(\alpha)$'s factor out in a
product of
$F_L^{(0)}(\alpha)$'s, where the $L$'s enter in the partition ${\cal P}$.
{}From this consideration, it follows that only the $F_N(\alpha)$'s should be
easily
accessible numerically. Note also that the regularisation due to the
lattice spacing necessarily alters the results, since
two
points cannot approach each other arbitrarily close
([8,9]).
In the case of a true brownian motion, when two points $M_I$ and $M_J$
come
close to each other, the accumulated angle $\theta_{IJ}$ has a non
negligible probability to become arbitrarily large (the typical
brownian path being non-differentiable,  $\theta_{IJ}$ follows a broad
law probability distribution).
Thus, when one tries to simulate brownian
paths
on a lattice, a correction is necessary
to take into account this characteristics of brownian
winding.
Each time two points $M_I$ and $M_J$ randomly walking on
the lattice are separated by only one
lattice
spacing, one adds to  $\theta_{IJ}$ a multiple
of
$2\pi$ which is drawn accordingly to a Cauchy law [9].
Typically, we have used random walks ranging from 50000 to 150000 steps.
The resulting numeric simulations for
the
$F_N^{(0)}(\alpha)$'s suggest the following simple formula
\begin{equation}
F_N^{(0)}(\alpha)=\prod_{k=1}^{N-1}(1-{N\over k}\alpha)
\end{equation}
valid for $0<\alpha<1$ (to obtain an expression in the whole interval
$[0,2]$ one uses
$F_N^{(0)}(\alpha)=F_N^{(0)}(-\alpha)=F_N^{(0)}(\alpha+2)$).

Eq.(19) correctly reproduces the $N=2$ and $N=3$ cases ;
it also has the right $\alpha=0$ and
$\alpha=1$ limits ; finally it
satisfies $F_N(1+\alpha)=(-1)^{N+1}F_N(\alpha)$.
Note that a similar formula has recently appeared in an a priori
different context [10],  the
one-dimensional Calogero model [11]. There, the
$F_N^{(0)}(\alpha)$'s, are derived  from the exact one dimensionnal $N$-body
spectrum (they are the coefficients appearing in the expansion of the
density in powers of the fugacity). This similarity is not a surprise :
it is
well-known that there is a correspondence [12] between the Calogero model
and the anyon model projected into the lowest Landau level of an
external
magnetic field. This might imply that when
anyons are projected into the Landau groundstate, mostly cycles of maximal
length contribute to the thermodynamics.

The expression obtained for $F_N^{(0)}(\alpha)$ is again quite simple. One
should be
able to find a way to support it by a pure analytical approach.
Let us concentrate on  the $N=3$ case.
As already stated, the antisymmetric part $F_{21}(\alpha)$ of
$Z_3(\alpha)$ is determined by
the known linear
states ;
its symmetric part is
\be Z_3(\alpha)+Z_3(1+\alpha)={2\over
3}F_3(\alpha)Z_1(3\beta) +{1\over
3}F_{111}(\alpha)Z_1(\beta)^3\ee
Could it be that $F_3$ is  also determined by the known part of the
$3$-anyon spectrum ?
Some simple considerations [13] on
$F_{111}$ will show
us the way.
The paths contributing to $F_{111}(\alpha)$ are simple one-particle
cycles
going back to their initial position at the time $\tau=\hbar\beta$.
The final configuration is
identical to the
initial one, thus   there is no need for quotienting  the configuration space
by the
permutation group $S_3$.
It follows that  $Z_1(\beta)^3\ ; F_{111}(\alpha)$ is simply the
partition function of a system of non-identical particles (Boltzmann
statistics) interacting
topologically \`a
la Aharonov-Bohm with coupling constant $\alpha$
now defined modulo 1 instead of
modulo 2 in the identical particle case.
In the case of $2$-anyon one indeed has that
\be Z_1(\beta)^2\ ;  F_{11}(\alpha)=Z_2(\alpha)+Z_2(\alpha+1)\ee
meaning that $F_{11}(\alpha)$ (for $2$ non identical particles) can be directly
obtained
from $Z_2(\alpha)$ (the partition function for 2 bosonic identical
particles)\footnote{In other words in $Z_2(\alpha)$ one has
summed on even angular momentum quantum numbers, and, in order to get the
complete spectrum, one simply shifts $\alpha\to \alpha+1$ to get the odd
angular momentum quantum numbers.}.
Thus one writes for $F_{111}$
\begin{equation}
Z_1(\beta)^3\ ;
F_{111}(\alpha)
=Z_3(\alpha,\alpha,\alpha)
+Z_3(\alpha+1,\alpha+1,\alpha+1)
+2 Z_3(mixed)
\end{equation}
with
\be Z_3(mixed)=Z_3(\alpha,\alpha,\alpha+1)+
Z_3(\alpha,\alpha+1,\alpha+1)\ee
where one has explicitly denoted the relative statistical
coefficient for each pair of particles (here $\alpha$ or $\alpha+1$).
In the case $\alpha=0$, this implies that the mixed representation of $S_3$ is
represented either by $(0,0,1)$ or by $(0,1,1)$, meaning in the first
case that the pairs
$\{1,2\}$ and $\{1,3\}$ are bosonic, and $\{2,3\}$ is fermionic, and in
the second case that the pair $\{1,2\}$ is bosonic, whereas $\{1,3\}$
and $\{2,3\}$ are fermionic.
The factor $2$ in front of $Z_3(mixed)$ is due to the fact that the symmetric
and
antisymmetric representations of $S_3$ are one-dimensional, but the mixed
one is
two-dimensional implying doubly degenerated states in the spectrum.

This in turn implies for $F_3(\alpha)$
\begin{equation}
Z_1(3\beta)F_3(\alpha)=Z_3(\alpha,\alpha,\alpha)+
Z_3(\alpha+1,\alpha+1,\alpha+1)-Z_3(mixed)
\end{equation}

This formula is analogous to (13), so one might ask about
its validity when one restricts oneself to linear states.
One has the generalized linear eigenstates [14]
\be \psi_{n,m_{IJ}}(r_{IJ},\theta_{IJ})=e^{-\beta
{\rho^2\over 2}}  \prod_{I<J}^{}e^{im_{IJ}\theta_{IJ}}
{r_{IJ}^{|m_{IJ}-\alpha_{IJ}|}
L_n^{N-2+\sum_{I<J}^{}|m_{IJ}-\alpha_{IJ}|}} (\beta \rho^2) \ee
\be E_{n,m_{IJ}}=\omega(2n+\sum_{I<J}|m_{IJ}-\alpha_{IJ}|+N-1)\ee
for a topological Hamiltonian where the Aharonov-Bohm coupling constants
$\alpha_{IJ}$ depend on the pair $\{I,J\}$ of particles coupled
($\rho^2=\sum_{I<J}r_{IJ}^2$).
Mixed linear states are obtained by trading
for each couple $\{I,J\}$ of particles
the
statistical parameter  $\alpha$  for
a
 statistical parameter  $\alpha_{IJ}$.
One can easily show that
the partition function for such mixed linear states is then directly obtained
from the
linear partition function $Z_3^{lin}(\alpha)$ by
${N(N-1)\over 2}\alpha \ ; \longrightarrow
\;\sum_{I<J}\alpha_{IJ}$.
For $N=3$
\begin{equation}
\begin{array}{ll}
Z_3^{ lin}(\alpha,\alpha,1+\alpha) &=
{\cosh\bigl[(2-3\alpha)\xi\bigr]\over   32\sinh^2{\xi\over 2}\sinh^2\xi
\sinh^2{ 3\xi\over 2}}\\
Z_3^{ lin}(\alpha,1+\alpha,1+\alpha) &=
{\cosh\bigl[(1-3\alpha)\xi\bigr]\over   32\sinh^2{\xi\over 2}\sinh^2\xi
\sinh^2{ 3\xi\over 2}} \end{array}
\end{equation}

Let us now consider (24) restricted to linear states only.
One finds that if the expression
\be Z_3^{lin}(\alpha,\alpha,\alpha)+Z_3^{lin}(\alpha+1,\alpha+1,\alpha+1)-
Z_3^{ lin}(\alpha,\alpha,1+\alpha)-Z_3^{lin}(\alpha,1+\alpha,1+\alpha)\ee
has not the correct finite limit $F_3^{(0)}(\alpha)$ in the thermodynamic
limit,
but on the contrary diverges, the left over divergence does not
depend on $\alpha$ and simply reads
$Z_1(\beta)Z_1(3\beta)$.
It is quite a striking fact that when one
adds $Z_1(\beta)Z_1(3\beta)$ to
$Z_3^{lin}(\alpha,\alpha,\alpha+1)+
Z_3^{lin}(\alpha,\alpha+1,\alpha+1)$ in (28)
one gets in the thermodynamic limit exactly the right $F_3^{(0)}(\alpha)$
as
given in (17).
Moreover, one can compute $F_3(\alpha)$ as given by (28) for an arbitrary
$\xi$
\begin{equation}
F_3(\alpha) =
{\sinh(2-3\alpha){\xi\over
2}\;\sinh(1-3\alpha){\xi\over 2}\over  \sinh{\xi\over 2}\;\sinh\xi}
\end{equation}
This expression generalizes quite nicely $F_2(\alpha)$ given in (8).

Similar considerations (albeit less transparent) can be made
for the $F_N^{(0)}$'s, $N>3$. One notes that  a
possible generalization of (29)
\be
F_N(\alpha)=\prod_{k=1}^{N-1}\Biggl({\sinh(k-N\alpha){\xi\over
2}\over
\sinh{k\xi\over 2}}\Biggr)\ee
would automatically satisfy the right thermodynamic limit
(19)
obtained in the numerical simulations.

To conclude this analysis, let us come back to the standard case
$\alpha=0$.
One knows that $Z_1^3(\beta)F_{111}(0)=Z_1^3(\beta)$ simply describes
$3$ independent bidimensionnal
oscillators ; one also has
\be Z_3(0,0,0)= {\cosh(3\xi)+ 2\cosh^2(\xi/2)\over 32\sinh^2{\xi\over
2}\sinh^2\xi \sinh^2{ 3\xi\over 2}}   \ee
\be Z_3(1,1,1)={1+ 2\cosh^2(\xi/2)\over 32\sinh^2{\xi\over
2}\sinh^2\xi \sinh^2{ 3\xi\over 2}}   \ee
One would like to find simple expressions for $Z_3(0,0,1)$ and
$Z_3(0,1,1)$.
It is clear from the bosonic (31) and fermionic (32) $3$-body partition
functions that the linear part of the spectrum contributes respectively
to $\cosh(3\xi)/d$ and $1/d$ (i.e. $\cosh(3-3\alpha)\xi/d $ for $\alpha=0,1$)
whereas the common $2\cosh^2(\xi/2)/d$
is built by the unknown
part of the spectrum ($d=
32\sinh^2{\xi\over
2}\sinh^2\xi \sinh^2{ 3\xi\over 2}$).
 It is quite natural to try for $Z_3(0,0,1)$ and
$Z_3(0,1,1)$
\be Z_3(0,0,1)\sim {\cosh(2\xi)+ 2\cosh^2(\xi/2)\over 32\sinh^2{\xi\over
2}\sinh^2\xi \sinh^2{ 3\xi\over 2}}   \ee
\be Z_3(0,1,1)\sim {\cosh\xi+ 2\cosh^2(\xi/2)\over 32\sinh^2{\xi\over
2}\sinh^2\xi \sinh^2{ 3\xi\over 2}}   \ee
and to check if (22) holds for $\alpha=0$. One finds again the same  left over
divergence $Z_1(\beta)Z_1(3\beta)$ as above
 (eventhough one gets the correct behavior $\sim{1\over
\xi^6}$ in the thermodynamic limit),  that should be added to
$Z_3(0,0,1)+Z_3(0,1,1)$ to get $Z_1(\beta)^3$.
It would certainly be interesting to understand more precisely, already
at the level of (22), the origin of this additional term, and also
in which way it splits between
$Z_3(0,0,1)$ and $Z_3(0,1,1)$, thus allowing a complete knowledge of the
$3$-body partition functions for the mixed eigenstates of $S_3$.

Acknowledgements : J. D. and S. O. acknowledge stimulating conversations with
J. Myrheim, and also for drawing our attention on [10].

Figure Captions :

A brownian closed path contributing to $F_3^{(0)}(\alpha)$. The points A,B
and C divide the curve into $3$ patches of equal length. The $3$ points
$M_1$, $M_2$ and $M_3$ span each of the patches with an equal speed
($M_1$ goes from A to B, etc...).

\vfill\eject

References:

[1] J. M. Leinaas and J. Myrheim, Nuovo Cimento B37, 1 (1977).

[2] R. Prange and S. Girvin, "The Quantum Hall Effect", Springer, New-York
(1987).

[3] J. Myrheim and K. Olaussen, Phys. Lett. B299, 267 (1993) ; J. Myrheim,
Thesis (Trondheim Preprint 1993).

[4] A. Comtet, J. Desbois and S. Ouvry, Journal of Physics A : Math. Gen.
23, 3699 (1990)

[5] D. P. Arovas, R. Schrieffer, F. Wilczek and A. Zee, Nucl. Phys. B251,117
(1985).

[6] for a review on linear states see A. Lerda, "Anyons"
Springer- Verlag (1992).

[7] D. Sen, Phys. Rev. Lett. 68, 2977 (1992).

[8] C. Belisle, Ann. Prob. 17, 1377 (1989).

[9] F. Spitzer, Trans. Ann. Math. Soc. 87, 187 (1958).

[10] S. B. Isakov, Int. Jour. Mod. Phys. A9, 2563 (1994) ;
Int. Jour. Mod. Phys. B8, 319 (1994).

[11] F. Calogero, J. Math. Phys. 10, 2191 (1969) ; 12, 419 (1971).

[12] L. Brink, T. H. Hansson, S. Konstein and M. A. Vasiliev,
Goteborg ITP-92-53 Preprint ;
  T. H. Hansson, J. M. Leinaas and J. Myrheim, Nucl.
   Phys. B384, 559 (1992) ;
 A. Dasnieres de Veigy and S. Ouvry, Phys. Rev. Lett. 72, 600 (1994) ;
Y-S Wu, Utah Preprint (1994) ; D. Bernard and Y-S Wu, Saclay Preprint
(1994) ; M. V. N. Murthy and R. Shankar, IMSc Preprint (1994).

[13] J. Myrheim, Private Communication.

[14] A. Dasnieres de Veigy and S. Ouvry, Phys. Lett. B307, 91 (1993).

 \end{document}